\documentclass[aps,prl,reprint,superscriptaddress,nobibnotes,nofootinbib]{revtex4-2}

\usepackage{amsmath, graphicx}

\usepackage{multirow}

\usepackage[T1]{fontenc}
\usepackage{placeins}

\usepackage{xurl}      % better URL line breaks
\usepackage{hyperref}  % clickable links
\usepackage[nameinlink]{cleveref}

\crefname{figure}{Fig.}{Figs.}
\Crefname{figure}{Figure}{Figures}
\crefname{table}{Table}{Tables}
\Crefname{table}{Table}{Tables}
\crefname{equation}{Eq.}{Eqs.}
\Crefname{equation}{Equation}{Equations}
\crefname{section}{Sec.}{Secs.}
\Crefname{section}{section}{Sections}

\crefrangeformat{equation}{Eqs.~(#3#1#4--#5#2#6)}
\Crefrangeformat{equation}{Equations~(#3#1#4--#5#2#6)}

\hypersetup{
    colorlinks=true,
    linkcolor=blue,
    filecolor=blue,      
    urlcolor=blue,
    citecolor=cyan,
    linktoc=all,
}

\makeatletter
\let\RT@origonecolumngrid\onecolumngrid
\let\RT@origtwocolumngrid\twocolumngrid

\renewcommand{\onecolumngrid}{%
  \RT@origonecolumngrid
  \onecolumn@grid@setup
  \let\set@footnotewidth\set@footnotewidth@one
  \let\compose@footnotes\compose@footnotes@one
}

\renewcommand{\twocolumngrid}{%
  \RT@origtwocolumngrid
  \twocolumn@grid@setup
  \let\set@footnotewidth\set@footnotewidth@two
  \let\compose@footnotes\compose@footnotes@two
}
\makeatother

\newcommand{\papertitle}{
Distinguishing between Direct and Parametric Driving in Nanomechanics Using a Vibrating Carbon Nanotube
}

\newcommand{\Pout}{\ensuremath{P_\text{out}}}
\newcommand{\Pdrive}{\ensuremath{P_\text{drive}}}
\newcommand{\Pprobe}{\ensuremath{P_\text{probe}}}
\newcommand{\PGamma}{\ensuremath{P_\Gamma}}
\newcommand{\Pgamma}{\ensuremath{P_\gamma}}

\newcommand{\fsb}{\ensuremath{f_\text{sb}}}
\newcommand{\fdrive}{\ensuremath{f_\text{drive}}}
\newcommand{\fprobe}{\ensuremath{f_\text{probe}}}

\newcommand{\omegam}{\ensuremath{\omega_\text{m}}}
\newcommand{\phim}{\phi_\text{m}}
\newcommand{\thetam}{\theta_\text{m}}
\newcommand{\omegadrive}{\ensuremath{\omega_\text{drive}}}
\newcommand{\phidrive}{\ensuremath{\phi_\text{drive}}}
\newcommand{\thetadrive}{\ensuremath{\theta
_\text{drive}}}
\newcommand{\omegaprobe}{\ensuremath{\omega_\text{probe}}}

\newcommand{\VGdrive}{\ensuremath{V_{\text{G},0}^\text{drive}}}
\newcommand{\Vsd}{\ensuremath{V_{\text{sd}}}}
\newcommand{\Vsdz}{\ensuremath{V_{\text{sd,0}}}}

\newcommand{\Vsdprobez}{\ensuremath{V_{\text{sd,p},0}}}

\newcommand{\Vsddrivez}{\ensuremath{V_{\text{sd,d},0}}}
\newcommand{\Vgdc}{\ensuremath{V_{\text{G,dc}}}}
\newcommand{\Vg}{\ensuremath{V_{\text{G}}}}

\newcommand{\kl}{\ensuremath{k_\ell}}
\newcommand{\klbar}{\ensuremath{\bar{k}_\ell}}
\newcommand{\Fdrive}{\ensuremath{F_\text{drive}}}
\newcommand{\Gammanl}{\ensuremath{\Gamma^\text{(nl)}}}

\makeatletter
\newcommand*{\currentname}{\@currentlabelname}
\newcommand{\speciallabel}[2]{% \speciallabel{<stuff>}{<label>}
  \edef\@currentlabel{#1}\label{#2}
}
\makeatother

\begin{document}

    \title{\papertitle}
    
    \author{Sam Dicker}
    \affiliation{Department of Physics, Lancaster University, Lancaster LA1 4YW, United Kingdom}
    
    \author{Patrick Steger}
    \affiliation{Department of Physics, Lancaster University, Lancaster LA1 4YW, United Kingdom}
    
    \author{Deepanjan Das}
    \altaffiliation[Present address: ]{Univ. Grenoble Alpes, CNRS, Grenoble INP,\\ Institut N\'eel, 38000 Grenoble, France}
    \affiliation{Department of Physics, Lancaster University, Lancaster LA1 4YW, United Kingdom}
    
    \author{Saba M. Khan}
    \altaffiliation[Present address: ]{Department of Applied Physics,\\ Aalto University, 02150 Espoo, Finland}
    \affiliation{Department of Physics, Lancaster University, Lancaster LA1 4YW, United Kingdom}
    
    \author{Edward A. Laird}
    \affiliation{Department of Physics, Lancaster University, Lancaster LA1 4YW, United Kingdom}

\date{\today}

\begin{abstract}
Parametric driving is a powerful route to amplification and nonlinear control in nanomechanical resonators, but its signatures can be ambiguous because standard dc electrical readout does not directly reveal the frequency of motion. Here we resolve this ambiguity by measuring the motional frequency of a vibrating carbon nanotube independently of the drive frequency. We operate the nanotube as an electromechanical mixer and detect microwave sidebands using a low-noise superconducting amplifier. This frequency-resolved readout distinguishes direct motion of the first overtone from parametric motion of the fundamental, even when the corresponding drive frequencies nearly coincide. The two mechanisms are further separated by their drive-power dependence. Beyond conventional parametric resonance at $2f_0$, we observe responses to driving at $3f_0$ and $4f_0$, consistent with high-order parametric excitation associated with nonlinear stiffness terms.\\ \\
Keywords: Nanomechanics, Carbon nanotubes, Parametric resonance
    % We distinguish directly driven from parametrically driven motion in a vibrating carbon nanotube by electrically measuring the frequency of motion independently of the drive frequency. The method up-converts mechanical motion to microwave sidebands, detected using a superconducting amplifier. The parametric responses are distinct in both motional frequency and dependence on the drive power. In addition to the well-known parametric response to driving at twice the fundamental frequency, we observe responses to driving at three and four times this frequency, arising from nonlinear stiffness.
\end{abstract}

\maketitle

\twocolumngrid

    A resonator can be driven directly, by applying a force near the resonant frequency, or parametrically, by modulating a system parameter at a frequency typically near twice the resonant frequency~\cite{Bachtold2022}.
    In electronics and optics, parametric driving is the basis for ultra-low-noise amplifiers~\cite{Aumentado2020}, noise squeezing, and the generation of entangled photons~\cite{Perelshtein2022}.
    In mechanical resonators, parametric coupling has been harnessed for amplification and thermomechanical noise squeezing~\cite{Rugar1991}, for studying synchronization~\cite{Shim2007}, and for detecting dielectric forces~\cite{Unterreithmeier2009}.
    Parametric nanomechanical experiments are challenging because the nonlinearity that gives rise to parametric responses is typically weak.
    However, carbon nanotube resonators have large and tunable nonlinearity because of the strong dependence of tension on gate voltage, which is further enhanced by Coulomb blockade~\cite{Steele2009,Lassagne2009}.
    This makes them ideal for studying parametric amplification~\cite{Eichler2011} and mode coupling~\cite{CastellanosGomez2012,Eichler2012}.
    Nevertheless, distinguishing directly driven from parametrically driven motion is difficult in the standard detection scheme, which measures the quasi-dc current through the moving nanotube~\cite{Sazonova2004}, especially because, for a tension-dominated string, the strongest parametric drive of the fundamental mode occurs near the frequency of the first overtone~\cite{Poot2012, Laird2012}.
    (We label the fundamental transverse mode as the zeroth mode and its first overtone as the first mode.)
    
    Here we show how to distinguish direct and parametric responses by directly measuring the frequency of motion in a nanomechanical carbon nanotube resonator.
    We do this by operating the device as a mixer and detecting the up-converted electromechanical sideband using a microwave amplification chain incorporating a near-quantum-limited superconducting amplifier.
    We identify the parametric responses when driving near twice a zeroth-mode frequency, and distinguish them from the directly driven motion in a first mode, for which the frequencies are nearby but distinct.
    This method allows us to identify high-order parametric resonances, not previously seen in nanomechanics, when driving at three and four times a zeroth-mode frequency.
    These are signatures of high-order nonlinearities of the nanotube stiffness.

    \begin{figure}
    \centering
    \includegraphics{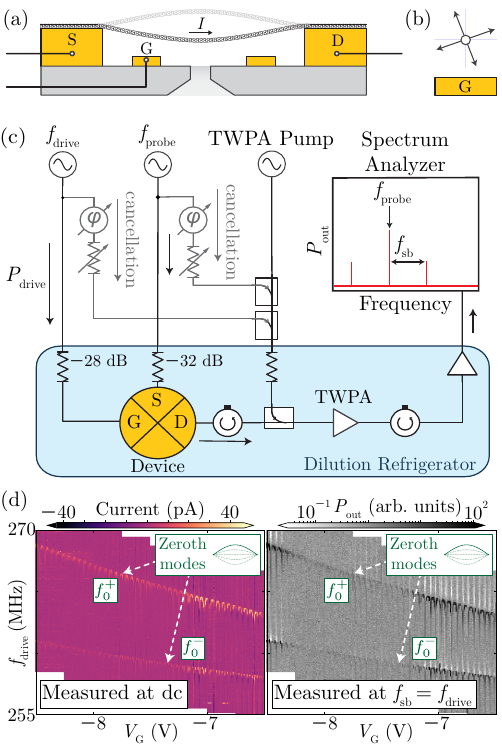}
    \caption{
    (a) Device schematic.
    The transverse flexural modes of the suspended nanotube are tuned using the dc part of the voltage on gate G, and excited by the ac part.
    To probe the motion, a dc + ac voltage is applied between source~S and drain D, inducing a current $I(t)$ through the device, which contains components that arise from electromechanical mixing.
    (b) Transverse displacement directions for a mode pair.
    (c) Simplified measurement circuit (dc path not shown; see \cref{app:FullCircuit} for full version).
    The nominal attenuation on the drive and probe paths is indicated.
    Sidebands in the output, appearing at frequency $\fsb$, arise when the nanotube moves at that frequency.
    (d) The zeroth (fundamental) mode pair of the nanotube, measured simultaneously in dc current (left) and ac power (right), as a function of dc gate voltage $V_{\text{G}}$ and gate drive frequency $\fdrive$.
    In the left panel, the average of each column was subtracted, and in the right panel, each column was normalized to increase clarity.
    The modes appear as a pair of lines at frequencies $f_0^\pm$, whose frequency depends on gate voltage because of electrostatic tension (which creates an overall slope) and Coulomb blockade (which creates a periodic modulation).}
    \label{fig:1}
    \end{figure} 

    % \section{Methods}
    Our resonator (\cref{fig:1}(a) and \cref{fig:Supp:1}(a)) consisted of a nanotube suspended between source and drain contacts~\cite{Huttel2008,Laird2012,Bachtold2022}, with flexural vibrations that are the mechanical modes studied.
    The device was cooled to $12$~mK and configured so that the central part of the nanotube formed a quantum dot in the Coulomb blockade regime, with Schottky tunnel barriers at the interfaces to the electrodes \cite{Laird2012}.
    To gate the quantum dot, and to tune and actuate mechanical motion, a dc voltage $\Vg$ combined with an oscillating voltage at frequency $\fdrive$ was applied to a gate electrode beneath the nanotube.
    $\Pdrive$ is the power of the drive signal injected into the dilution refrigerator.
    For all measurements, a 9~mV dc bias was applied between source and drain.
    
    To investigate the mechanical response of this device, it was configured for simultaneous dc and ac measurements (\cref{fig:1}(c) and \cref{app:FullCircuit}).
    The dc measurements revealed the average change in current through the device in response to a drive; the ac measurements revealed the frequency at which the nanotube was moving.
        
    The dc part of the current $I$ through the nanotube was measured using a room-temperature transimpedance amplifier followed by a digital multimeter.
    For ac measurements, the device was operated as a mixer.
    With a probe tone at frequency $\fprobe=5$~GHz added to the source voltage, the motion of the nanotube modulates its conductance at the mechanical frequency, giving rise to sidebands in the current that are offset by $\fsb$ from the probe tone.
    The ac components of the current were amplified using a superconducting traveling-wave parametric amplifier (TWPA)~\cite{Perelshtein2022,Rej2025}, followed by semiconductor amplifiers.
    To avoid saturation and unintentional mixing in the TWPA, cancellation tones were injected along separate paths to interfere destructively with signals from $\fdrive$ and $\fprobe$ that inevitably coupled through the device.
    Finally, the amplified signal was measured using a spectrum analyzer whose resolution bandwidth was set to $30$~Hz.
    The mechanical resonances are evident in both the dc and ac response, and form a ladder of flexural mode pairs $f_0^\pm, f_1^\pm, \cdots$.
    The subscripts label the order of the corresponding beam mode, with each pair being split~\cite{Moser2013} into an upper and a lower branch (labeled by superscripts $+$ and $-$) because the nanotube can vibrate in two directions (\cref{fig:1}(b)).
    \Cref{fig:1}(d) shows the two branches of the zeroth mode pair measured in dc and in ac.

    \begin{figure}
    \centering
    \includegraphics{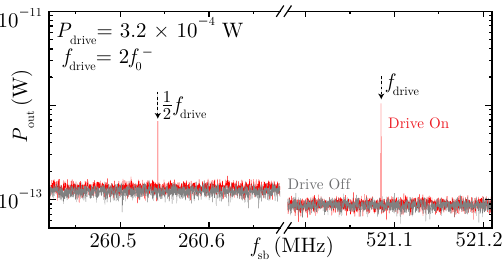}
    \caption{Output power as a function of frequency when the device is driven at $\fdrive=521.09$~MHz at a gate voltage $-8514$~mV.
    The peak at $\fsb = \fdrive/2$ shows that the lower branch of the zeroth mode pair ($f_0^-\approx 260.54$~MHz) is being excited by a drive at $2f_0^-$, indicating a parametric mechanism.
    A peak also appears at $\fdrive$, which results from a combination of pure electrical mixing (i.e., not caused by mechanical motion), leakage signals into the TWPA, and the nonlinear electrical response to motion at $\fdrive/2$.
    As expected, no peaks are seen when the drive is off.}
    \label{fig:2}
    \end{figure}

    \begin{figure*}[ht]
    \centering
    \includegraphics[width=\textwidth]{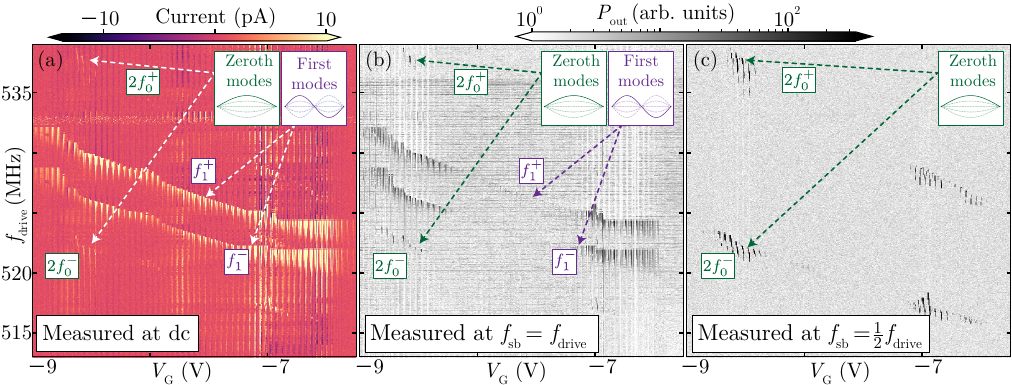}
    \caption{
    Measurements around the first mode pair $f_1^\pm$ (i.e. the first overtone) using dc current (a) and ac detection (b, c).
    Panel (b), measured simultaneously with (a), shows the ac component at the drive frequency; panel (c) shows the component at half the drive frequency.
    Both the $f_0^\pm$ modes and the $f_1^\pm$ modes are visible in this figure, with the $f_1^\pm$ pair being directly driven (and therefore moving at $\fdrive$) and the $f_0^\pm$ pair being parametrically driven at $2f_0^\pm$ (and therefore moving at $\fdrive/2$).
    Consistent with this picture, the directly driven modes appear strongly both at dc (a) and at the drive frequency~(b); however, the directly driven modes do not appear in (c), because motion at $\fdrive$ generates no signal at $\fdrive/2$.
    For all panels, the drive power injected into the refrigerator was $1 \times 10^{-4}$~W, and the probe tone was injected at $3.16 \times 10^{-5}$~W.
    To make the resonances clear, in (a), the average of each column was subtracted and in (b, c) each column was normalized.
    }
    \label{fig:3}
    \end{figure*}

    % \section{Results}
    \Cref{fig:2} demonstrates parametric resonance detected using the ac measurement.
    The drive was applied at $\fdrive=521.09$~MHz, approximately twice the mechanical frequency $f_0^-$.
    This induced sidebands at both $\fprobe \pm \fdrive$ and $\fprobe \pm \frac{1}{2}\fdrive$.
    The former can arise simply from mixing between the two input frequencies, due to electrical nonlinearity.
    However, the latter cannot arise from electrical mixing alone; instead, it is a signature of parametric response in a zeroth mode to a drive at twice its frequency of motion.

    \begin{figure}
    \centering
    \includegraphics[width=9cm]{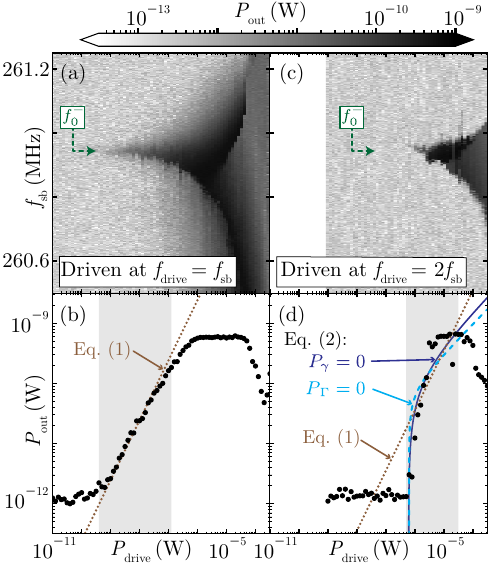}
    \caption
    {Direct and parametric driving of the same resonance.
    (a) Output power, measured at $\fsb=\fdrive$, as a function of $\Pdrive$ and $\fdrive$ as $\fdrive$ is scanned to directly drive the $f_0^-$ mode.
    (b) Points: maximum of each column in (a). Line: fit to \cref{eq:main:direct_drive_power}, as expected for a directly driven resonator with linear displacement-to-current transduction.
    (c) The same measurement as in (a), but now driving at twice the frequency of (a) and measuring at $\fsb=\fdrive/2$. This is parametric excitation.
    (d) Points: maximum of each column in (c).
    Lines: Fits to \cref{eq:main:direct_drive_power} (dotted) and \cref{eq:main:parametric_drive_power} (solid/dashed).
    Agreement with \cref{eq:main:parametric_drive_power} is better, confirming that the mechanism is parametric.
    (Shading in panels (b) and (d) indicates fitting ranges.)}
    \label{fig:4}
    \end{figure}

    To further investigate the parametric response, \cref{fig:3} shows simultaneous dc and ac measurements as a function of gate voltage $V_{\text{G}}$ and drive frequency $\fdrive$, when the device was driven around the first mode pair $f_1^\pm$ (i.e. the first overtone; see insets).
    As expected for a beam in the high-tension limit, the $f_1^\pm$ mode pair is at approximately twice the frequency of the $f_0^\pm$ pair.
    The dc measurement (\cref{fig:3}(a)) shows two pairs of resonances, a bright inner pair and a dim outer pair, with the dim outer pair appearing at drive frequencies equal to twice the fundamental mode frequencies $f_0^\pm$ measured in \cref{fig:1}(d).
    The ac measurement shows different behavior depending on whether the signal is measured at $\fsb = \fdrive$ (\cref{fig:3}(b)) or at $\fsb=\fdrive/2$ (\cref{fig:3}(c)).
    When $\fsb = \fdrive$, the same modes appear as in the dc measurement; however, when $\fsb=\fdrive/2$, only the outer pair appears.
    
    These results can be explained by identifying the inner pair of resonances as due to direct driving of the $f_1^\pm$ modes, and the outer pair as parametric driving of the $f_0^\pm$ modes.
    In the dc measurement (\cref{fig:3}(a)), both types of response are visible, because both types of motion change the average current.
    Both responses also appear in \cref{fig:3}(b).
    Direct driving of the first mode induces motion, and hence a current component, at the drive frequency.
    Parametric driving induces motion at half the drive frequency in the zeroth mode; the nonlinear conductance of the device generates an electrical overtone that appears as the dim outer pair~(\cref{app:high_order}).
    However, only the outer pair of modes shows a response at $\fsb = \fdrive/2$ (\cref{fig:3}(c)); this is the signature that these modes are being parametrically driven, leading to motion (and therefore a changing current) at $\fdrive/2$.
    Importantly, these observations cannot be explained by ordinary electrical mixing, since that generates responses only at the drive frequency or an integer multiple thereof; in contrast, \cref{fig:3}(c) shows a response at half the drive frequency.%$\fsb = \fdrive/2$.
    
    We can further distinguish direct from parametric response by comparing how different resonances respond to different drive powers (\cref{fig:4}).  
    When the detected sideband amplitude is proportional to the displacement, a directly driven linear resonator has~\cite{Meerwaldt2012}
    \begin{equation}
        \Pout = a\Pdrive,
        \label{eq:main:direct_drive_power}
    \end{equation}
    where $\Pout$ is the output power measured at $\fsb$, $a$ is a proportionality constant, and $\Pdrive$ is the drive power.
    However, a parametrically driven resonance is expected to obey~\cite{Bachtold2022}
    \begin{equation}
        \Pout = \text{max}\bigl\{0, \, b(\Pdrive - \Pgamma)^{1/2} - \PGamma\bigr\},
        \label{eq:main:parametric_drive_power}
    \end{equation}
    where $b$, $\Pgamma$, and $\PGamma$ are fit parameters related to the Duffing nonlinearity and nonlinear damping (\cref{app:PowerRelation}).
    Thus a parametric resonance should show a turn-on with driving power.
    
    \Cref{fig:4} compares the ac response of the $f_0^-$ mode under direct driving (panels (a, b)) and parametric driving (panels (c, d)).
    Panel (a) shows the output power when this mode was driven close to its mode frequency (i.e. with $\fdrive\approx f_0^-$) and measured at $\fsb = \fdrive$.
    As expected, the response became stronger as the drive power was increased.
    Panel (b) plots the strength of the on-resonance response, extracted by taking the maximum of each column in (a), as a function of $\Pdrive$.
    Fitting this to \cref{eq:main:direct_drive_power} confirms that over a wide range of drive power, the output is proportional to the drive.

    Panel (c) is the corresponding measurement when the same mode was driven near twice its mode frequency (i.e. with $\fdrive\approx 2f_0^-$) but measured at $\fsb = \fdrive/2$.
    In this panel, the resonance appears abruptly as a function of drive power.
    Panel (d) plots the maximum of each column.
    The data are now poorly fit by~\cref{eq:main:direct_drive_power} but well fit by \cref{eq:main:parametric_drive_power}, confirming that panel (c) corresponds to parametric driving and panel (a) corresponds to direct driving.
    Two versions of the fit are shown, one with $\Pgamma$ set to zero (corresponding to zero Duffing parameter) and one with $\PGamma$ set to zero (corresponding to zero nonlinear damping); the former gives a better fit, suggesting that the saturation of the response is dominated by nonlinear damping rather than by Duffing nonlinearity (although Duffing nonlinearity is clearly present, as shown by the asymmetric line shape in \cref{fig:4}(a)).
    At low $\Pdrive$, the fits deviate from the data because the signal was below the noise floor; at high $\Pdrive$, the signal is smaller than predicted by \crefrange{eq:main:direct_drive_power}{eq:main:parametric_drive_power}, which is consistent with nonlinear electrical response.
    
\begin{figure*}
    \centering
        \includegraphics{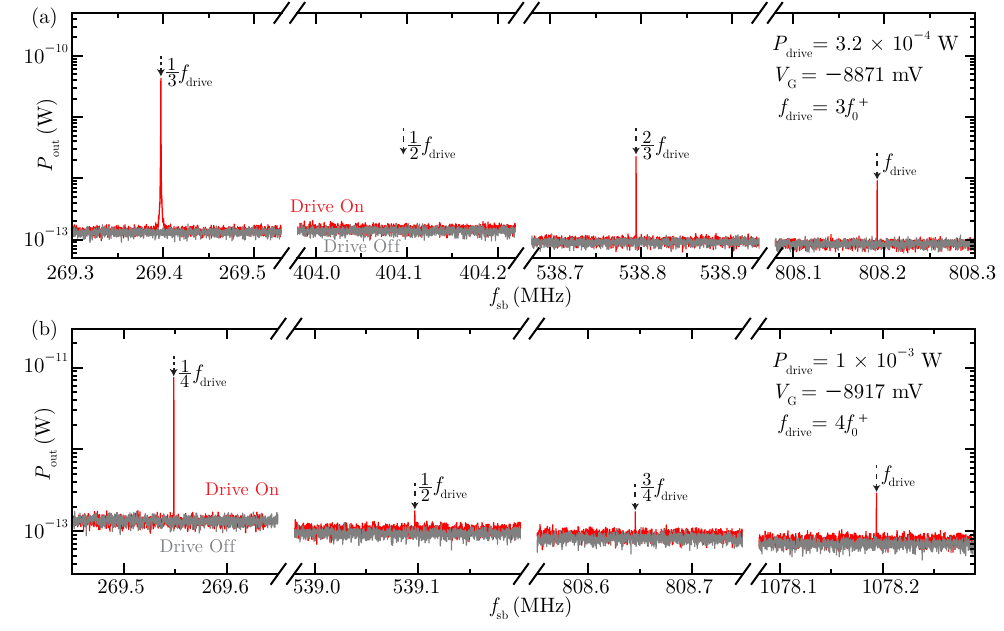}
    \caption{High-order parametric driving.
    (a) Parametric driving at $\fdrive = 3f_0^+$.
    The fact that the response appears at $\fdrive/3$ and is absent at $\fdrive/2$ shows that this is parametric driving via the nonlinear $k_2$ stiffness term (see text).
    The peak at $2\fdrive/3$ is due to the motion at $\fdrive/3$ generating a higher electrical overtone due to nonlinear conductance; the peak at $\fdrive$ is a combination of this effect with leakage and pure electrical mixing.
    (b) Parametric driving at $\fdrive = 4f_0^+$ exciting motion at $\fdrive/4$, with the corresponding electrical overtones also visible.
    }
    \label{fig:5}
    \end{figure*}

    The data so far show the signatures of conventional parametric resonance, which occurs when the resonator is driven close to twice its zeroth-mode frequency~\cite{Rugar1991,Eichler2011}.
    However, we also observe responses when driving at higher multiples of this frequency, not explained by the conventional theory that considers only the linear stiffness.
    \Cref{fig:5} shows this effect.
    
    \Cref{fig:5}(a) shows the response of the device when driven at approximately three times the zeroth-mode frequency (upper branch).
    In conventional parametric resonance, the corresponding motion can occur at frequencies~\cite{Kovacic2018}
    \begin{equation}
        \fsb=\frac{n\fdrive}{2},
        \label{eq:main:allowed_frequencies}
    \end{equation}
    where $n$ is a positive integer. 
    However, \cref{fig:5}(a) shows responses at $\fdrive/3$ and $2\fdrive/3$, neither of which is allowed by \cref{eq:main:allowed_frequencies}; the first allowed frequency, which is $\fdrive/2$, shows no response.
    \Cref{fig:5}(b) shows similar behavior when driving at four times the zeroth-mode frequency; once again, there are responses at non-allowed frequencies $\fdrive/4$ and $3\fdrive/4$, as well as at the allowed frequency $\fdrive/2$.

    To explain these results, we extend the theory of parametric resonance by including high-order nonlinearities of the spring constant. 
    An analogous electrical high-order parametric driving has previously been predicted~\cite{Zhang2017} and measured in a superconducting resonator at three, four, and five times its zeroth-mode frequency~\cite{Svensson2018}.
    The equation of motion for a nonlinear resonator with one relevant mode at $f_0$ is
        \begin{equation}
        \ddot{u}+\frac{2\pi f_0}{Q}\dot{u} +  \frac{1}{M} \sum_{\ell=1}^\infty \kl u^\ell=0
            \label{eq:main:nonlinear_resonator}
        \end{equation}
    where $u(t)$ is displacement as a function of time $t$, $Q$ the quality factor, $M$ the mass, $\kl$ the $\ell$th stiffness coefficient of the resonator, and $2\pi f_0 = \sqrt{k_1/M}$.
    Parametric driving can be induced by modulating the stiffness coefficients about their average values:
    \begin{align}
        \kl(t)  &= \klbar+\delta\kl(t) \\
                &= \klbar \, [1 + h_\ell \cos (2\pi\fdrive t)],
        \label{eq:main:nonlinear_driving}
    \end{align}
    where $h_\ell$ are the modulation amplitudes.
    Given an initial displacement $u(t)\propto \cos (2\pi f_0t)$, driving can occur when $\delta\kl(t) \, u^\ell(t)\propto \cos (2\pi \fdrive t)\, \cos^\ell (2\pi f_0 t)$ contains a term at the mechanical frequency $f_0$.
    This condition is satisfied when $\fdrive = (\ell+1)f_0$ for some integer $\ell$.

    Conventional parametric response results from modulating the $\ell=1$ term in \cref{eq:main:nonlinear_resonator}, leading to a response at half the drive frequency when $\fdrive \approx 2 f_0$.
    However, \cref{fig:5}(a) is instead explained as a consequence of the $\ell=2$ term in \cref{eq:main:nonlinear_resonator}, which allows for parametric driving at $\fdrive \approx 3f_0$.
    The device moves at $f_0$, which leads to a spectral peak at $f_0$ because the conductance is changing at that frequency, and smaller peaks at $2f_0$ and $3f_0$ consistent with nonlinear conductance producing electrical overtones.
    Similarly, \cref{fig:5}(b) is evidence of the $\ell=3$ term, which gives a parametric response from driving at $\fdrive\approx 4f_0$.
    
    % \section{Conclusion}
    We have integrated a carbon nanotube resonator with a near-quantum-limited TWPA, thus combining two technologies for high force sensitivity and low-noise electrical amplification.
    The resulting sensitivity and bandwidth allow us to distinguish unambiguously between direct and parametric driving by measuring the frequency of the motion directly.
    Identifying the underlying mechanical spectrum of a resonator in the presence of parametric resonances, which cannot be cleanly distinguished in dc current alone, enables better characterization of the mode structure and may be useful for nanomechanical mass sensors~\cite{Hanay2015, Chaste2012}, in which knowing the shape of different modes is necessary to infer the added mass. 
    Our technique also offers a way to detect high-order nonlinearities in the stiffness coefficient and to drive at three or four times the frequency of motion.

    In superconducting electronics, parametric coupling has allowed for ultra-sensitive amplifiers and control of single-photon states.
    In nanomechanics, this ability to control and characterize parametric effects is a step toward mechanical amplifiers~\cite{Olkhovets2001,Aleman2011,Eichler2011} capable of high force and mass sensitivity.
    Ultimately, it may allow for detection of phase superpositions in flexural modes of the entire nanotube~\cite{Zhang2017}.

    \section*{Data and code availability}
    
    A full data repository including all figures, Python scripts, and raw data is available at 
    \href{https://zenodo.org/records/20447040?preview=1&token=eyJhbGciOiJIUzUxMiJ9.eyJpZCI6IjNhMDgwOWY1LWMwZTItNGY1NC1iMzJiLTA2MmI5MTkxMzEzYyIsImRhdGEiOnt9LCJyYW5kb20iOiIxNjNlNGIyNGZhNmU3NTA4ZGE2N2MzODVkOWZkYmJkYiJ9.mc48U_n_soC7sWlO-goorL06QPsjSEYfy43gCeSCmNOhrCqRfd7oDJEAuxUx3ZnRkEqoSNoRJ3by6BrwVc6r5w}{https://doi.org/10.5281/zenodo.20447040}.
   
    \section*{Acknowledgements}
    The authors acknowledge the team at VTT Technical Research Centre of Finland Ltd. for providing the JTWPA: Joonas Govenius, Leif Grönberg, Robab Najafi Jabdaraghi, Janne Lehtinen, Mika Prunnila, and Visa Vesterinen.
    The authors thank A. Eichler for suggestions.
    The authors acknowledge the EU (grants 818751 and 824109), STFC (ST/W006502/1), and EPSRC (EP/Z534250/1). 

    \bibliographystyle{apsrev4-2}
    \bibliography{references}
    \vfill

 \clearpage
\onecolumngrid
\raggedbottom

%%%%%%%%%%
% Supplementary
%%%%%%%%%%

\setcounter{page}{1}
\setcounter{equation}{0}
\setcounter{figure}{0}
\setcounter{table}{0}
\setcounter{section}{0}
\setcounter{subsection}{0}

\setcounter{secnumdepth}{2}

\renewcommand{\thepage}{S\arabic{page}}
\renewcommand{\theequation}{S\arabic{equation}}
\renewcommand{\thefigure}{S\arabic{figure}}
\renewcommand{\thetable}{S\arabic{table}}
\renewcommand{\thesection}{\Alph{section}}
\renewcommand{\thesubsection}{\thesection.\arabic{subsection}}

% Unique hyperref anchors in the Supplementary
\renewcommand{\theHequation}{S\arabic{equation}}
\renewcommand{\theHfigure}{S\arabic{figure}}
\renewcommand{\theHtable}{S\arabic{table}}
\renewcommand{\theHsection}{S\arabic{section}}
\renewcommand{\theHsubsection}{S\arabic{section}.\arabic{subsection}}

\phantomsection
\pdfbookmark[1]{Supporting Information}{supplementary}

\thispagestyle{plain}

\begin{center}
\vspace*{-1.0ex}
{\small\scshape Supporting Information for}\\[0.8ex]
{\large\bfseries \papertitle\par}
\vspace{0.8ex}
{\normalsize Sam Dicker, Patrick Steger, Deepanjan Das, Saba M. Khan, and Edward A. Laird\par}
\vspace{1.2ex}
\rule{0.82\textwidth}{0.4pt}
\end{center}

\vspace{0.8ex}

\begin{center}
\begin{minipage}{0.72\textwidth}
\small
\setlength{\parindent}{0pt}
\renewcommand{\arraystretch}{1.18}

\textbf{Contents}\par
\vspace{0.5ex}

\begin{tabular}{@{}r@{\hspace{0.7em}}p{0.84\linewidth}@{}}
\hyperref[app:DeviceStructure]{A.} &
  \hyperref[app:DeviceStructure]{Device fabrication and characterization} \\

\hyperref[app:SupportingData]{B.} &
  \hyperref[app:SupportingData]{Supporting data} \\

\hyperref[app:FullCircuit]{C.} &
  \hyperref[app:FullCircuit]{Details of the measurement circuit} \\

\hyperref[app:DrivingNanotube]{D.} &
  \hyperref[app:DrivingNanotube]{Origin of electromechanical signals} \\

\hyperref[app:high_order]{E.} &
  \hyperref[app:high_order]{High-order parametric excitations} \\

\hyperref[app:PowerRelation]{F.} &
  \hyperref[app:PowerRelation]{Derivation of Eqs.~(1) and (2)}
\end{tabular}
\end{minipage}
\end{center}

\vspace{1.5ex}
    
    \section{Device fabrication and characterization}\label{app:DeviceStructure}

    \begin{figure}[b]
    \centering
        \includegraphics[width = \textwidth]{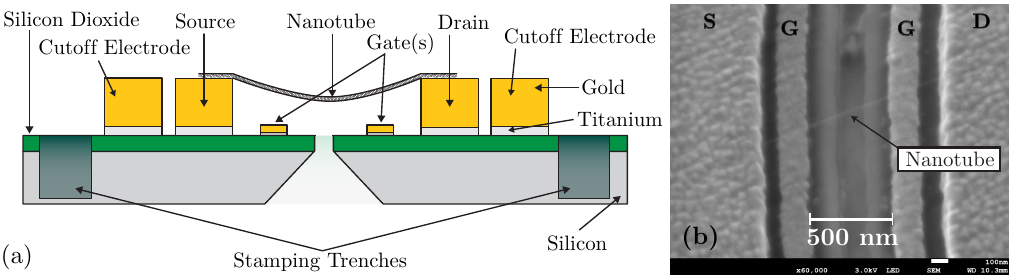}
        \caption{(a) Side-view schematic of the device.
        The cutoff electrodes and the trenches are used for stamping fabrication; the source, drain, and gate electrodes are used for the experiment.
        (b) SEM image of the device. S, D, and G indicate source, drain, and gate respectively.}
        \label{fig:Supp:1}
    \end{figure}

    \Cref{fig:Supp:1}(a) shows a schematic of our device.
    The nanotube is suspended between source and drain electrodes above two gate electrodes; cutoff electrodes and trenches were used during the fabrication process.
    The device was fabricated by a stamping technique~\cite{Wu2010, Waissman2013, Blien2018}, described in detail in Ref.~\cite{Steger2025}.
    To create the device chip, electrodes were patterned on Si/SiO$_2$ using a combination of electron-beam lithography and photolithography, thermal evaporation, and lift-off.
    To enable later stamping, two trenches were etched using a thick etch mask and reactive ion etching.
    A through-chip aperture was also etched beneath the device to allow for future deposition of material onto the center of the nanotube, but this was not done for this experiment.
    
    Separately, the nanotube was grown by chemical vapor deposition between the tines of a silicon fork.
    Catalyst particles on the fork tips had been formed by thermally evaporating a nominally $\sim 1$~nm thick Co layer.
    The fork was loaded into a modified probe station and stamped down onto a previously wire-bonded device by lowering the tips into the stamping trenches.
    During this process, the conductance between the two cutoff electrodes was monitored to detect when the nanotube had made contact; once this was detected, the nanotube was electrically burned between the cutoff electrodes and the source/drain electrodes by supplying a large current.
    The fork was then retracted, leaving the suspended segment behind.
    The completed device is shown in \cref{fig:Supp:1}(b).
    The right-hand gate was found to be disconnected, and was therefore not used.
       
\begin{figure}
    \centering
    \includegraphics[width = \textwidth]{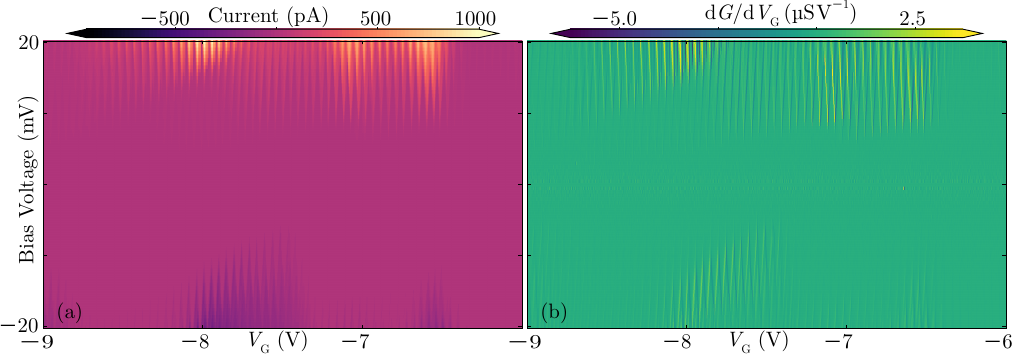}
    \caption{(a) Stability diagram of the nanotube measured without drive or probe tones, showing Coulomb diamonds. 
    (b) Derivative of the conductance, calculated from (a). The ac signal is expected to be proportional to this derivative (see \cref{eq:Supp:Iac1}), since $dG/du \propto dG/dV_\mathrm{G}$.}
    \label{fig:Supp:2}
    \end{figure}

    The device was loaded into a carrier PCB assembly, inside the puck of an Oxford Instruments Triton dilution refrigerator.  
    \Cref{fig:Supp:2} shows the source-drain current as a function of bias voltage and gate voltage.
    The regular Coulomb diamonds, evident both in the current and in the derivative of the conductance, show that the device acts as a single-electron transistor.

    \section{Supporting data}\label{app:SupportingData}
    Here we present further characterization of the mechanical resonances.
    
    \subsection{Characterization of the zeroth-mode pair}
\begin{figure}[b]
    \centering
    \includegraphics[width = \textwidth]{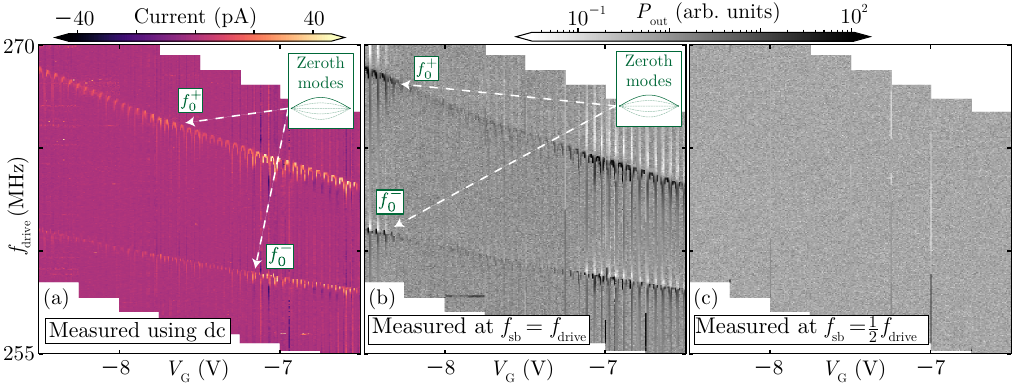}
    \caption{
    Equivalent of \Cref{fig:3} but with driving near the zeroth mode pair measured in dc current (a), in ac at the drive frequency (b), and in ac at half the drive frequency (c).
    The absence of a signal in (c) confirms that this mode is directly rather than parametrically driven.
    Panels (a) and (b) are the same as \cref{fig:1}(d) in the main text.
    In all panels, $\Pdrive=1 \times 10^{-4}$~W and $\Pprobe = 3.16 \times 10^{-5}$~W.
    In (a), the average of each column was subtracted, and in (b, c), each column was normalized to increase clarity.
    }
    \label{fig:Supp:3}
    \end{figure}

    \Cref{fig:Supp:3} shows full characterization near the zeroth-mode frequencies, as \cref{fig:3} in the main text does for the first mode pair.
    As expected, driving the resonator at its mechanical frequency generates signals both at dc (panel (a)) and in ac at the drive frequency (panel (b)).
    However, unlike in \cref{fig:3}, there is no ac signal at half the drive frequency (panel (c)).
    This confirms that this mode pair is being directly rather than parametrically driven, and since there are no observable modes at lower frequency we identify it as the zeroth mode pair.
    
    \subsection{\texorpdfstring{Parametric signal as a function of $\fdrive$ and $\fsb$}{}}
    
\begin{figure}
    \centering
    \includegraphics[width = \textwidth]{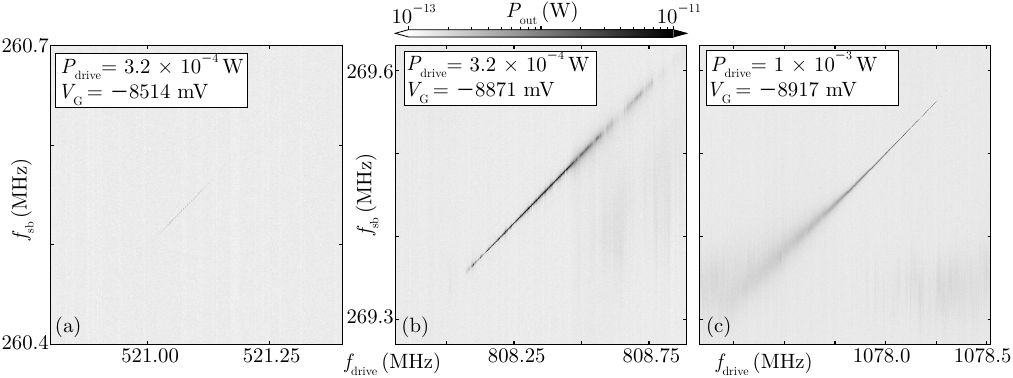}
    \caption{
    (a) Parametric responses from the lower branch of the zeroth-mode, seen at $\fsb \approx f_0^{-}$ while scanning $\fdrive$ across $2f_0$. 
    (b) and (c) Corresponding responses seen at $\fsb \approx f_0^{+}$ while sweeping $\fdrive$ across $3f_0$ and $4f_0$.
    }
    \label{fig:Supp:4}
    \end{figure}

    To further confirm the mechanical nature of the parametric response, \Cref{fig:Supp:4} shows the output spectrum as a function of drive frequency. 
    Panel (a) shows the lower branch of the zeroth-mode pair when driven near twice its frequency; panels (b) and (c) show the upper branch when driven near three and four times its frequency, respectively.
    As expected, the response frequency tracks the drive frequency; furthermore, parametric response occurs only over a range of frequencies centered on the mode frequency and corresponding approximately to the mechanical linewidth.

    %\FloatBarrier
    \section{Details of the Measurement Circuit}\label{app:FullCircuit}
    
\begin{figure*}
    \centering
        \includegraphics[width = \textwidth]{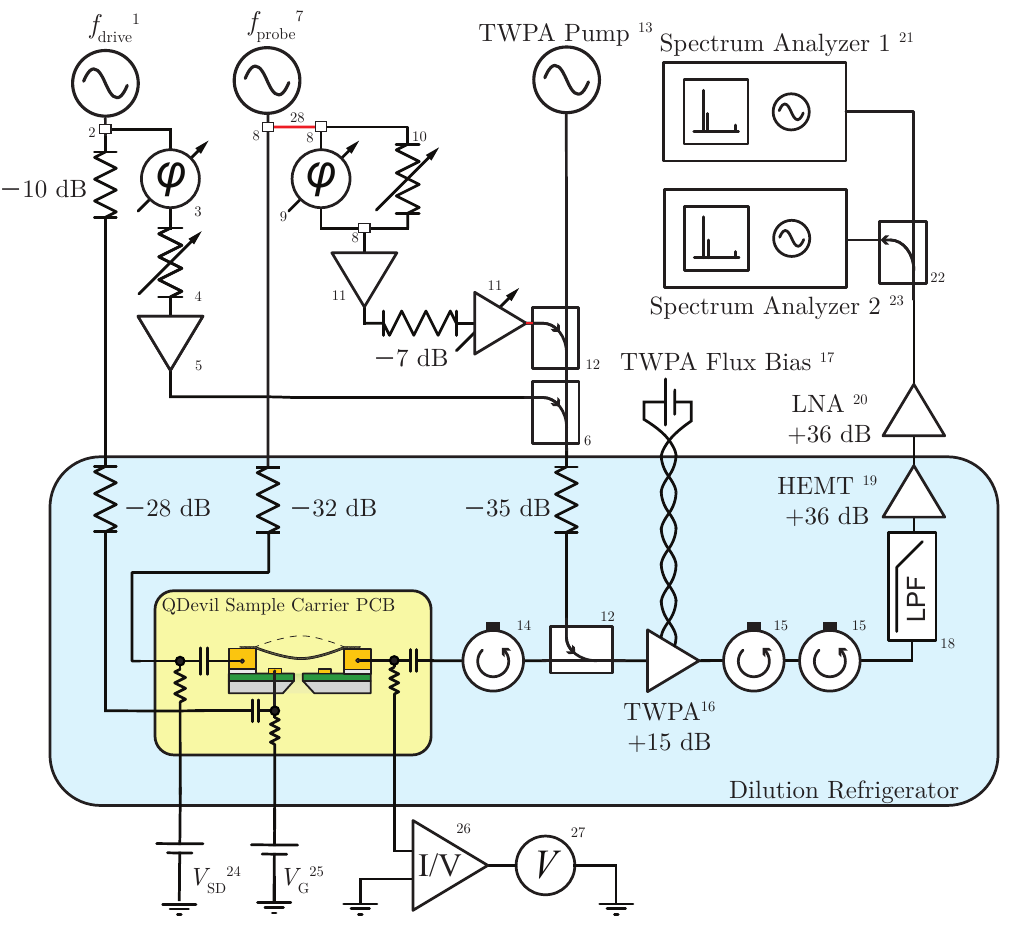}
        \caption{Detailed measurement circuit.
        The components are listed in~\cref{tab:Supp:PartList}.
        The cryogenic attenuation values shown represent totals over attenuators placed at different thermal stages of the refrigerator.
        }
        \label{fig:Supp:5}
    \end{figure*}

    \Cref{fig:Supp:5} and \cref{tab:Supp:PartList} detail the full circuitry for simultaneous dc and ac measurements.
    The nanotube was driven using an rf signal generator at frequency $\fdrive$ applied to the gate, and probed using a tone at $\fprobe=5$~GHz applied to the source.
    Both signals were applied via attenuators for cooling and bias tees for adding dc voltages.
    The current through the device was separated using a bias tee into dc and ac measurement chains.
    The dc measurement chain was an I/V converter followed by a digital voltmeter.
    The ac chain consisted of a Josephson TWPA (JTWPA) at 12~mK followed by a HEMT amplifier at 4~K and a room-temperature low-noise amplifier (LNA); a series of isolators in the chain protect the device from the amplifier pump tone and thermal noise.
    The JTWPA was a pre-commercial AI-TWPA-C from Arctic Instruments, consisting of a periodic array of Josephson junction-based meta-atoms embedded in a thin-film waveguide structure~\cite{Perelshtein2022}.
    The JTWPA supports broadband non-degenerate three-wave mixing when biased with an external dc flux and driven by a microwave pump tone.
    It was pumped at $12$~GHz and $5.5\times 10^{-4}$~W, with a flux bias applied from a battery-powered source.
    The flux bias was set using a current of $819~\mu$A, which over a range from $4$ to $8$~GHz was found to give a gain and noise temperature of $\sim 15$~dB and $\sim 500 $~mK respectively, approximately three times the standard quantum limit.
    To stabilize the gain of the LNA, which was affected by fluctuations in the room temperature (despite air-conditioning) a \href{https://www.thingiverse.com/thing:6579572}{heat sink and a fan} were used to keep it at a stable temperature.
    Following amplification, the output power was measured using Spectrum Analyzer 1, which was set to acquire near $\fprobe+\fsb$.

    % Although the measurement frequency was well away from $\fdrive$ and $\fprobe$, we found that the gain and noise of the TWPA were still degraded by leakage of these tones past the nanotube and into its input.
    We found that the gain and noise of the TWPA were degraded by leakage of $\fdrive$ and $\fprobe$ past the nanotube and into its input.
    To prevent this, we added a pair of cancellation lines in order to suppress these signals via destructive interference at the TWPA input.
    Each line allowed for tunable amplitude and phase, and successful cancellation was monitored using Spectrum Analyzer 2 to ensure that the leaked power through the ac chain was close to zero.

    To set the attenuation and phase in the drive cancellation path, before each sweep, $\fdrive$ was varied over the entire measurement range, and at each setting, the setpoints for phase and attenuation were adjusted to give near-zero leakage power.
    After fitting to functions of a suitable shape, this calibration data can then be used to calculate appropriate parameters for the cancellation path for all setpoints of the experiment.

    In the probe cancellation path, even though $\fprobe$ was not changed, this procedure was found to be inadequate, because the transmission was not stable over the course of a sweep.
    We therefore implemented the following dynamic procedure for ensuring the cancellation condition was met.
    First, the phase shifter in the path was set to approximately the correct value, determined by minimizing the leakage signal at Spectrum Analyzer 2.
    Then, a variable attenuator, in parallel with the phase shifter, was used for fine adjustment of the phase, followed by fine adjustment of the amplitude by adjusting an amplifier gain. (This was achieved by tuning the amplifier supply voltage.)
    In order to maintain cancellation, the variable attenuator and the variable amplifier were adjusted up to approximately once per second during the sweep, using a computer-controlled feedback loop to minimize the leakage to Spectrum Analyzer 2.

    The largest contribution to the fluctuations in the feedback paths came from the varying temperature of the lab, which affected the phase through the cables and amplifiers.
    To suppress cable fluctuations, we used nominally matched pairs of cables, and where that was not possible, phase-stable cables that minimize the effect of Teflon temperature dependence (the ``Teflon knee''~\cite{Carlisle2017}).
    To reduce amplifier fluctuations, the components in the microwave cancellation path were placed in an insulated box.
     
    \begin{table*}
    \caption{Component list for the detailed measurement circuit (\cref{fig:Supp:5}).}
    \label{tab:Supp:PartList}
    \centering
    \small
    \setlength{\tabcolsep}{3pt}
    \renewcommand{\arraystretch}{1.1}
    
    \resizebox{\textwidth}{!}{%
    \begin{tabular}{|l|l|l|l|l|l|l|}
    \hline
    \begin{tabular}[c]{@{}l@{}}Component \\ Number\end{tabular} & Component and Model & Powered by &  & \begin{tabular}[c]{@{}l@{}}Component \\ Number\end{tabular} & Component and Model & Powered by \\ \hline
    1 & \begin{tabular}[c]{@{}l@{}}RF Signal Generator\\ Rohde \& Schwarz SMBV100A\end{tabular} & Mains power &  & 15 & \begin{tabular}[c]{@{}l@{}}Circulator\\ LNF-CICIC4-12A\end{tabular} & N/A \\ \hline
    2 & \begin{tabular}[c]{@{}l@{}}Power Splitter\\ Mini-Circuits ZESC-2-11+\end{tabular} & N/A &  & 16 & \begin{tabular}[c]{@{}l@{}}TWPA \\ Arctic Instruments AI-TWPA-C (pre-commercial)\end{tabular}  & \begin{tabular}[c]{@{}l@{}}Flux Bias$^{17}$\\ and TWPA Pump$^{13}$\end{tabular} \\ \hline
    3 & Custom Phase Shifter~\cite{Yin2014}& Keithley 2450 SMU &  & 17 & \begin{tabular}[c]{@{}l@{}}Battery Powered Voltage \\ Source at 819~$\mu$A\end{tabular} & 9~V Alkaline Battery \\ \hline
    4 & \begin{tabular}[c]{@{}l@{}}Digital Attenuator\\ Vaunix LDA-102\end{tabular} & USB &  & 18 & \begin{tabular}[c]{@{}l@{}}Low Pass Filter\\ Marki Microwave FLP-0750 7.5~GHz low-pass\end{tabular} & N/A \\ \hline
    5 & \begin{tabular}[c]{@{}l@{}}RF Amplifier\\ Mini-Circuits ZX60-33-LN-S+\end{tabular} & \begin{tabular}[c]{@{}l@{}}Keysight E3630A \\ dc Power Supply\end{tabular} &  & 19 & \begin{tabular}[c]{@{}l@{}}HEMT \\ LNF-LNC0.3 14A\end{tabular} & Mains power \\ \hline
    6 & \begin{tabular}[c]{@{}l@{}}Directional Coupler\\ Pasternack PE2210-10\end{tabular} & N/A &  & 20 & \begin{tabular}[c]{@{}l@{}}Low Noise Amplifier\\ Narda LNA-30-04001200-15-10P\end{tabular} & \begin{tabular}[c]{@{}l@{}}Keysight E3630A \\ dc Power Supply\end{tabular} \\ \hline
    7 & \begin{tabular}[c]{@{}l@{}}MW Signal Generator\\ Rigol DSG3136B\end{tabular} & Mains power &  & 21 & \begin{tabular}[c]{@{}l@{}}Spectrum Analyzer 1\\ Rohde \& Schwarz FSV3044\end{tabular} & Mains power \\ \hline
    8 & \begin{tabular}[c]{@{}l@{}}MW Power Splitter\\ Mini-Circuits ZFRSC-183-S+\end{tabular} & N/A &  & 22 & \begin{tabular}[c]{@{}l@{}}Directional Coupler\\ Pasternack PE2210-20\end{tabular} & N/A \\ \hline
    9 & \begin{tabular}[c]{@{}l@{}}Digital Phase Shifter\\ Vaunix LPS-802\end{tabular} & USB &  & 23 & \begin{tabular}[c]{@{}l@{}}Spectrum Analyzer 2\\ Rigol DSA875\end{tabular} & Mains power \\ \hline
    10 & \begin{tabular}[c]{@{}l@{}}Digital Attenuator\\ Vaunix LDA-802EH\end{tabular} & USB &  & 24 & \begin{tabular}[c]{@{}l@{}}Bias Voltage Supply\\ QuTech IVVI-DAC2 - D5 and S3b modules\end{tabular} & Lead Acid Battery \\ \hline
    \multirow{2}{*}{11} & \multirow{2}{*}{\begin{tabular}[c]{@{}l@{}}MW Amplifier\\ Mini-Circuits ZX60-24A-S+\end{tabular}} & \begin{tabular}[c]{@{}l@{}}Keysight E3630A \\ dc Power Supply\end{tabular} &  & 25 & \begin{tabular}[c]{@{}l@{}}Gate Voltage Supply\\ QuTech IVVI-DAC2 - D5 and S1f modules\end{tabular} & Lead Acid Battery \\ \cline{3-7} 
     &  & Keithley 2450 SMU &  & 26 & \begin{tabular}[c]{@{}l@{}}Transimpedance Amplifier and Output\\ QuTech IVVI-DAC2 - M1b and M0b modules\end{tabular} & Lead Acid Battery \\ \hline
    12 & \begin{tabular}[c]{@{}l@{}}Directional Coupler\\ Krytar 120420\end{tabular} & N/A &  & 27 & \begin{tabular}[c]{@{}l@{}}Digital Multimeter\\ Keithley 2100\end{tabular} & Mains power \\ \hline
    13 & \begin{tabular}[c]{@{}l@{}}TWPA Pump\\ Windfreak SynthHD\end{tabular} & Mains power &  & 28 & \begin{tabular}[c]{@{}l@{}}Phase Insensitive Cables\\ \href{https://www.rfone.cn/products/tp220}{TP220}\end{tabular} & N/A \\ \hline
    14 & \begin{tabular}[c]{@{}l@{}}Circulator\\ LNF-CIC4-12A\end{tabular} & N/A &  &  &  &  \\ \hline
    \end{tabular}
    }
    \end{table*}

    \section{Origin of Electromechanical Signals}\label{app:DrivingNanotube}

    In this section we explain the origin of the electromechanical signals appearing in the dc and ac measurements.
    The instantaneous current through the nanotube (approximating linear conductance $G$) is
    \begin{equation}
        I(t)=\Vsd(t)\,G(u(t))
        \label{eq:Supp:ohm}
    \end{equation}
    where the source-drain bias is 
    \begin{equation}
    \Vsd(t) = \Vsdz + \Vsdprobez\,\cos(\omegaprobe t) + \Vsddrivez\,\cos\thetadrive
    \label{eq:Supp:Vsd}
    \end{equation}
    This bias includes the applied dc bias $\Vsdz$, the applied probe voltage $\Vsdprobez \cos(\omegaprobe t)$, and a presumed contribution $\Vsddrivez\cos\thetadrive$ due to asymmetric parasitic coupling of the drive frequency to the contact electrodes, where
    \begin{equation}
        \thetadrive     \equiv \omegadrive t + \phidrive
    \end{equation}
    and $\phidrive$ is the phase of the parasitic coupling.
    (As usual, $\omegaprobe\equiv 2\pi \fprobe$, and similarly for other frequencies.)

    The conductance is $G(u(t))$, which depends on the displacement because the nanotube is a single-electron transistor~\cite{Chiu2008}, can be expanded as
    \begin{align}
        G(t)&= G_{0} + G'u(t) + \frac{1}{2}G''u^2(t) + \ldots\\
            & = G_{0} + G'A\cos\thetam + \frac{1}{4}G''A^2(1+\cos2\thetam) + \ldots
    \label{eq:Supp:G}
    \end{align}
    where the displacement is
    \begin{equation}
        u(t) = A \cos \thetam.
    \end{equation}
    Here $A$ is the displacement amplitude, 
    \begin{align}
        \thetam         &\equiv \omegam t + \phim
    \end{align}
    and $\phim$ is the phase of the motion.

    The experiment is sensitive to contributions of $I(t)$ at frequencies near dc or near $\omegaprobe$.
    Substituting \cref{eq:Supp:Vsd} and \cref{eq:Supp:G} into \cref{eq:Supp:ohm} and keeping only those terms gives
    \begin{equation}
    \begin{alignedat}{2}
        I(t) = &~\rlap{\text{(terms independent of displacement)}}\\ 
             &+ \frac{\Vsdz G'' A^2}{4}\\
             &+ \Vsdprobez G' A \cos \omegaprobe t \,\cos \thetam
             &&+ \frac{\Vsdprobez G'' A^2}{4} \cos \omegaprobe t \,(1 + \cos 2\thetam) \\
             &+ \Vsddrivez G' A \cos \thetadrive \,\cos \thetam
             &&+ \frac{\Vsddrivez G'' A^2}{4}  \cos \thetadrive \,\cos 2\thetam \\
             &+\cdots
    \end{alignedat}
    \end{equation}
    The displacement therefore leads to a dc contribution
    \begin{equation}
        I_\text{dc} = \frac{\Vsdz G'' A^2}{4}
        + \Vsddrivez G' A \langle \cos \thetadrive \cos \thetam \rangle
        + \frac{\Vsddrivez G'' A^2}{4}  \langle \cos \thetadrive \cos 2\thetam \rangle
        \label{eq:Supp:Idc}
    \end{equation}
    and ac contributions
    \begin{align}
        I_{\omegaprobe\pm\omegam}(t) &= 
        \Vsdprobez G' A \cos \omegaprobe t \cos \thetam 
        \label{eq:Supp:Iac1}\\
    I_{\omegaprobe\pm 2\omegam} (t) &= \frac{\Vsdprobez G'' A^2}{4}
         \cos \omegaprobe t \cos 2\thetam
         \label{eq:Supp:Iac2}
    \end{align}
    at $\omegaprobe\pm\omegam$, $\omegaprobe\pm2\omegam$, and so on.
    These are the signals amplified by the ac measurement chain and appearing in the measured spectrum as in \cref{fig:1}.
    Higher-order sidebands arise from higher terms in \cref{eq:Supp:G} not modeled here.

    The dc measurement is sensitive to the contribution in \cref{eq:Supp:Idc}.
    The ac measurement is sensitive to the contributions in \crefrange{eq:Supp:Iac1}{eq:Supp:Iac2}, which appear as sidebands to the probe tone in the output.
           
    \section{High-order parametric excitations}\label{app:high_order}

    \subsection{Derivation of the high-order parametric excitation frequencies}

    Here we derive the drive frequencies at which parametric resonance is excited by modulating nonlinear spring coefficients.
    We model a resonator subject to a restoring force
    \begin{equation}
        F(u) = -k(u) \, u
    \end{equation}
    where $u(t)$ is the displacement as a function of time $t$ and $k(u) \equiv \sum_{\ell=1}^{\infty}\kl\,u^{\ell-1}$ is the nonlinear spring coefficient.
    Under parametric driving, the equation of motion, obtained by combining \crefrange{eq:main:nonlinear_resonator}{eq:main:nonlinear_driving}, is
    \begin{equation}
        \frac{d^2u(t)}{dt^2}+\frac{\omega_0}{Q}\frac{du(t)}{dt} +      
        \frac{1}{M} \sum_{\ell=1}^\infty \klbar \, (1 + h_\ell \cos \omegadrive t) \,\bigl[u(t)\bigr]^\ell=0.
        \label{eq:supp:EOM}
    \end{equation}
    As in the main text, $\omega_0 \equiv \sqrt{\overline{k_1}/M}$ is the mode frequency, $Q$ the quality factor, $M$ the mass, $\omegadrive>0$ the drive frequency, and $h_\ell$ a set of parameters that characterize the strength of the drive.

    We identify those drive frequencies at which a small initial displacement is amplified.
    To do this, we first rewrite \cref{eq:supp:EOM} in terms of two small parameters $g$ and $h$, by defining
    \begin{align}
    \frac{\klbar}{M}     &\equiv g\,\alpha_\ell,   \qquad \ell=2,3,4, \ldots \\[4pt]
    \frac{\klbar h_\ell}{M} &\equiv h\,\beta_\ell,    \qquad \ell=1,2,3, \ldots.
    \end{align}
    For simplicity, we ignore damping by setting $Q \rightarrow \infty$.
    \Cref{eq:supp:EOM} then becomes
        \begin{equation}
        \frac{d^2u(t)}{dt^2}+      
         \omega_0^2 u(t) + g\sum_{\ell=2}^\infty \alpha_\ell \, \bigl[u(t)\bigr]^\ell+ h\sum_{\ell=1}^{\infty}\beta_\ell \, \bigl[u(t)\bigr]^\ell \cos \omegadrive t = 0.
        \label{eq:supp:EOM2}
    \end{equation}

    We now do perturbation theory~\cite{Nayfeh1995} in $h$, by writing
    \begin{equation}
        u(t) \approx u_0(t) + h u_h(t)
        \label{eq:Supp:perturbation_theory}
    \end{equation}
    and requiring \cref{eq:supp:EOM2} to be satisfied at zeroth and first order in $h$. This gives:
    \begin{align}
        \frac{d^2u_0(t)}{dt^2} + \omega_0^2 u_0(t) + g\sum_{\ell=2}^\infty \alpha_\ell \,[u_0(t)]^\ell
        &=
        0 
        \label{eq:Supp:EOMh_zero}
        \\
        \frac{d^2u_h(t)}{dt^2} + \omega_0^2 u_h(t)
        &=
        - g\sum_{\ell=1}^{\infty} (\ell+1) \, \alpha_{\ell+1} \,[u_0(t)]^\ell u_h(t) 
        - \sum_{\ell=1}^\infty \beta_\ell \,[u_0(t)]^\ell \cos \omegadrive t.
        \label{eq:Supp:EOMh_h}
    \end{align}
    \Cref{eq:Supp:EOMh_zero} describes unforced motion in an anharmonic potential.
    The displacement is assumed to be small, so the anharmonicity can be neglected and the general solution is
    \begin{equation}
        u_0(t) = A_0 \cos(\omega_0 t + \phi_0)
        \label{eq:Supp:u0}
    \end{equation}
    where $A_0$ and $\phi_0$ are arbitrary constants.
    The left-hand side of \Cref{eq:Supp:EOMh_h} describes harmonic motion of $u_h(t)$, with driving on the right-hand side.
    The amplitude will increase if the right-hand side 
    contains secular terms, i.e., components at the frequency $\omega_0$.
    These can arise from the anharmonicity alone (via the first sum) or from the driving (via the second sum).
    Since we want to know the effect of the drive, we neglect the former (which is also justified in the limit of weak anharmonicity, i.e., $g \rightarrow 0$).
    Substituting \cref{eq:Supp:u0} into \cref{eq:Supp:EOMh_h} then gives
    \begin{align}   
        \frac{d^2u_h(t)}{dt^2} + \omega_0^2 u_h(t)
        &= - \sum_{\ell=1}^\infty \beta_\ell \, A_0^\ell \, \cos^\ell (\omega_0 t + \phi_0)\, \cos \omegadrive t
        \\
        &=  - \sum_{\ell=1}^\infty \beta_\ell \, A_0^\ell \, \sum_{m=0}^{\lfloor \ell/2 \rfloor}
        \frac{2-\delta_{2m,\ell}}{2^\ell}\frac{\ell!}{m!(\ell-m)!}\,
        \cos[(\ell-2m)(\omega_0 t + \phi_0)] \, \cos \omegadrive t
        \label{eq:Supp:EOM_h}
    \end{align}
    where the second line follows from a trigonometric identity.
    
    To see when the right-hand side of \cref{eq:Supp:EOM_h} contains terms at $\omega_0$, expand
    \begin{equation}
        \cos\bigl[(\ell-2m)(\omega_0 t+\phi_0)\bigr]\cos\omegadrive t
        =\frac{\cos\bigl[(\ell-2m)(\omega_0 t+\phi_0)+\omegadrive t\bigr]
        +\cos\bigl[(\ell-2m)(\omega_0 t+\phi_0)-\omegadrive t\bigr]}{2}.
    \end{equation}
    This contains components at $\omega_0$ when
    \begin{equation}
        \bigl|(\ell-2m)\omega_0\pm\omegadrive \bigr| =\omega_0.
        \label{eq:Supp:parametric_condition}
    \end{equation}
    For each value of $\ell$ represented in \cref{eq:Supp:EOM_h}, the corresponding parametric driving frequencies are those that satisfy \cref{eq:Supp:parametric_condition} for some value of $m$, i.e.,
    \begin{equation}
        \frac{\omegadrive}{\omega_0}=\bigl|\ell-2m\pm 1\bigr|,
        \qquad
        m=0,1,2,\dots,\left\lfloor\frac{\ell}{2}\right\rfloor,
    \end{equation}
    or more simply (remembering that $\omegadrive$ is restricted to be positive):
    \begin{equation}
        \frac{\omegadrive}{\omega_0}
        =
        \begin{cases}
            1, 3, \, \dots, \ell+1, & \ell\ \text{even},\\[4pt]
            2, 4, \, \dots,\,\ell+1, & \ell\ \text{odd}.
        \end{cases}
        \label{eq:Supp:parametric_frequencies}
    \end{equation}
    As expected, the first-order case $\ell=1$, which corresponds to conventional parametric driving through modulation of the linear spring constant, leads to the well-known primary parametric resonance \cite{Kovacic2018}, visible e.g., in \cref{fig:2}, at
    \begin{equation}
        \omegadrive = 2 \omega_0,
        \label{eq:supp:2omega}
    \end{equation}
    for which the motion is at frequency $\omegadrive/2$.
    (This is the $n=1$ case of \cref{eq:main:allowed_frequencies}.)
    
    High-order parametric driving occurs when the modulation has $\ell \geq 2$.
    Parametric resonance at $\omegadrive/\omega_0 = 3$ or 4, as in \cref{fig:5}, arises from the $\ell=2$ and $\ell=3$ contributions to~\cref{eq:supp:EOM}.
    The resonances observed in \cref{fig:5} are the final elements in the sequence \cref{eq:Supp:parametric_frequencies}, i.e., $\omegadrive=(\ell+1)\omega_0$.

    The $n>1$ cases of \cref{eq:main:allowed_frequencies} arise when $\ell=1$ and the perturbation expansion of \cref{eq:Supp:perturbation_theory} is expanded to higher order~\cite{Nayfeh1995, Kovacic2018}.
    This expansion shows that the parametric drive frequency can be smaller than \cref{eq:supp:2omega}, but never larger as in \cref{eq:Supp:parametric_frequencies}.
    \Cref{fig:5} is therefore unambiguous evidence of high-order parametric resonance, i.e., modulation of a $\kl$ term with $\ell>1$.
    \footnote{It has been pointed out by Zhang~\emph{et al.}~\cite{Zhang2017} for the case $\ell=2$ that a variant of this mechanism is for direct driving at $(\ell+1)\omega_0$ to couple via the $\klbar$ term in \cref{eq:supp:EOM}.
    In this case, the direct drive at $(\ell+1)\omega_0$ is equivalent to a modulation of the $\kl$ stiffness coefficient.
    For simplicity (and because our data do not distinguish between the two contributions) we have not included in our model this contribution to the high-order parametric driving, although it is likely to be present.}

    \subsection{Excluding other causes of subharmonics}
    While our data clearly do not arise from generation of harmonics, there are in principle sources of subharmonics that at first sight resemble a parametric response.
    Here we consider these and say why they do not explain our data.

    \begin{figure*}
    \centering
    \includegraphics{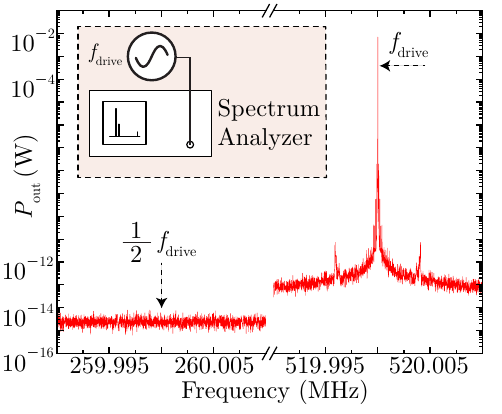}
    \caption{Spectrum of the rf signal generator when $\fdrive = 520$~MHz, measured as in the inset. No spectral impurity appears at $\fdrive/2$; the weak impurity sidebands offset by $\sim 32$~kHz do not mimic parametric resonances.}
    \label{fig:Supp:6}
    \end{figure*}
    
    Firstly, a spectrum similar to~\cref{fig:2} would arise if the drive tone at $\fdrive$ contained a spectral impurity at $\fdrive/2$.
    To exclude this possibility, \cref{fig:Supp:6} shows the spectrum of the $\fdrive$ signal generator set to a similar frequency.
    Any impurity at $\fdrive/2$ is smaller than the main tone by at least 12 orders of magnitude, and therefore cannot explain the $\fdrive/2$ peak in~\cref{fig:2}, which is smaller by less than 1 order of magnitude.
    The probe signal may also contain spectral impurities, but these would be easily identifiable in \cref{fig:Supp:3} or \cref{fig:3} as having no dependence on gate voltage.

    Secondly, period-doubling or period-multiplication can accompany a stable resonance of a directly driven anharmonic resonator~\cite{Holmes1981}.
    In this situation, which is distinct from parametric resonance, the resonator moves primarily at the driving frequency $\fdrive$, but also with components at $\fdrive/2$ or lower fractions.
    While superficially this resembles the data of \cref{fig:2} and \cref{fig:5}, it does not explain our data.
    If it did, the peaks in \cref{fig:3}(c) would track the positions of the first mode pair (i.e. the strong peaks in \cref{fig:3}(a,b)), since those are the resonances which are directly driven in the range shown.
    Instead, they track the zeroth modes, confirming that these are the modes being excited.
    
    \section{Derivation of equations (1) and (2)}\label{app:PowerRelation}
        Here we derive \crefrange{eq:main:direct_drive_power}{eq:main:parametric_drive_power}, used to fit \cref{fig:4}.
        The driving force, arising from the gate voltage, is
        \begin{align}
            F(t) &= \frac{1}{2} C' (\Vgdc + \VGdrive\cos \omegadrive t)^{2}\\
                    &\propto \Vgdc^{2} + 2\Vgdc\VGdrive\cos \omegadrive t+\frac{1}{2}(\VGdrive)^{2}(1+\cos 2\omegadrive t) \\
                    &\approx F_0 + \Fdrive \cos \omegadrive t
                    \label{eq:ESForceAC}
        \end{align}
        where $C'$ is the derivative of the gate capacitance with respect to $u$, and $\Vgdc$ and $\VGdrive$ characterize the static and oscillating parts of the gate voltage, respectively.
        The amplitude of the driving force at the drive frequency is therefore
        \begin{equation}
            \Fdrive = C' \Vgdc \VGdrive \propto \sqrt{\Pdrive}.
        \end{equation}
        In a directly driven linear resonator, the resulting amplitude is
        \begin{equation}
            A = \frac{\Fdrive}{M\Gamma \omega_{0}}
            \label{eq:DirectAmplitudeToDrive2}
        \end{equation}
        where $\Gamma \equiv \omega_0/Q$ is the linear damping coefficient.
        The output power is measured on the upper sideband and therefore, by \cref{eq:Supp:Iac1},
        \begin{align}
            \Pout   &\propto \langle I^{2}_{\omegaprobe + \omegam} \rangle \propto A^2 \propto \Fdrive^2 \label{eq:supp:Pu} \\
                    &\propto \Pdrive
        \end{align}
        which is \cref{eq:main:direct_drive_power}.

        In parametric driving, the relationship between driving strength and amplitude is more complicated.
        In the simplest case, which presumably applies in \cref{fig:4}, driving is via modulation at $2\omega_0$ of the linear $k_1$ component of the stiffness.
        The force component that this contributes at $\omega_0$ is proportional to both the modulation and the amplitude of the oscillations, i.e., is an effective negative damping.
        If the driven resonator were otherwise linear, there would be two possible outcomes: for weak parametric driving the intrinsic damping would overwhelm the drive and the displacement would be zero; for strong driving the displacement would increase without bound.
        The existence of stable parametric oscillations, as in \cref{fig:4}(c,d), shows that either the stiffness or the damping must be nonlinear.

        The amplitude in this case is calculated in Ref.~\cite{Bachtold2022}.
        If the equation of motion is
        \begin{equation}
            \frac{d^2u(t)}{dt^2} + \left(\Gamma + \Gammanl u^2(t)\right) \frac{du(t)}{dt} + \omega_0^2 u(t) + \gamma u^3(t) = \frac{h_1\bar{k}_1}{M} u(t) \cos 2\omega_0 t
        \end{equation}
        where $\Gammanl$ and $\gamma \equiv k_3/M$ are the nonlinear damping and Duffing coefficients, respectively, then the stable non-zero amplitude $A$ satisfies
        \begin{equation}
            \left(\Gamma + \frac{\Gammanl}{4} A^2\right)^2
            + \left(\frac{3\gamma}{4\omega_0}A^2\right)^2
            =
            \left(\frac{h_1\bar{k}_1}{2M\omega_0}\right)^2
            \label{eq:Supp:bachtold}
        \end{equation}
        The solution of \cref{eq:Supp:bachtold} with real $A$ is
        \begin{equation}
            A^2 = 
            \sqrt{
            \frac{h_1^2\bar{k}_1^2}{M^2\omega_0^2 K}
            -\left(\frac{3\Gamma\gamma}{\omega_0 K}\right)^2
            }
            -\frac{\Gamma\Gammanl}{K}
        \label{eq:Supp:u0solution}
        \end{equation}
        where
        \begin{equation}
            K \equiv \left(\frac{\Gammanl}{2}\right)^2 + \left(\frac{3\gamma}{2\omega_0}\right)^2.
        \end{equation}
        Using
        \begin{align}
            A^2   &\propto \Pout \\
            h_1^2   &\propto \Pdrive
        \end{align}
        and combining the various proportionality constants and parameters in \cref{eq:Supp:u0solution} gives
        \begin{equation}
            \Pout = b(\Pdrive-\Pgamma)^{1/2}-\PGamma
        \end{equation}
        which is \cref{eq:main:parametric_drive_power}.
        Here 
        \begin{align}
            \Pgamma &\propto \gamma^2/K^2 \\
            \PGamma &\propto \Gammanl/K
        \end{align}
        are parameters expressing the relative importance of nonlinear damping and nonlinear restoring force in determining the stable amplitude, and $b$ is an overall proportionality constant, parameterizing among other things the overall strength of the parametric drive.

        \Cref{fig:4}(b, d) was fitted using least-squares regression, with each datapoint weighted as if its uncertainty is proportional to its power, and with the fitting range judged to be where the resonance peak is visible but not distorted.
        Panel (d) shows two fits, one with $\Pgamma$ constrained to be zero,
        giving $b = (1.6 \pm 0.3) \times 10^{-7}~\text{W}^{1/2}$ and $P_{\Gamma} = (1.2 \pm0.3) \times 10^{-10}$~W,
        and one with $\PGamma$ constrained to be zero,
        giving  $b= (1.0 \pm 0.2) \times 10^{-7}~\text{W}^{1/2}$ and $\Pgamma=(6.30 \pm 0.02) \times 10^{-7}$~W.
        A fit with both $\Pgamma$ and $\PGamma$ as free parameters, but constrained to be non-negative, converges to the fit with $\Pgamma=0$.

\end{document}